\documentclass[apl,amsmath,reprint,floatfix,a4paper]{revtex4-1}

\usepackage[T1]{fontenc}
\usepackage{etex}
\usepackage{geometry}
\usepackage{color,tabularx,colortbl}

\usepackage{setspace}
\usepackage{pgfplots}
\usepackage{booktabs}
\usepackage{multirow}
\usepackage{tikz, tikz-3dplot}
\tikzset{>=latex}
\usepackage[percent]{overpic}
\pgfplotsset{compat=1.13}
\usepgfplotslibrary{external} 
\usepgfplotslibrary{fillbetween}
\usepgfplotslibrary{colorbrewer}
\usetikzlibrary{pgfplots.groupplots,arrows,decorations.markings}
\usetikzlibrary{shapes.geometric,calc,patterns}
\usepackage[strict]{changepage}

\usepackage{sidecap}
\sidecaptionvpos{figure}{t}

\usepackage{todonotes}
\makeatletter
\renewcommand{\todo}[2][]{\tikzexternaldisable\@todo[#1]{#2}\tikzexternalenable}
\renewcommand{\missingfigure}[2][]{\tikzexternaldisable\@missingfigure[#1]{#2}\tikzexternalenable}
\makeatother

\usepackage{dcolumn}
\newcolumntype{d}[1]{D{.}{.}{#1}}

\pgfmathsetmacro{\sigmaGaussFWHMlarge}{1.69864360058}
\pgfmathsetmacro{\sigmaGaussFWHMsmall}{0.849321800288}
\pgfmathsetmacro{\Tmin}{270}
\pgfmathsetmacro{\TClarge}{644.326893845}
\pgfmathsetmacro{\TCsmall}{536.939078204}

\pgfmathsetmacro{\Tfreeze}{364}
\pgfmathsetmacro{\Tjump}{442.335666891}
\pgfmathsetmacro{\tfreeze}{0.483233608052}
\pgfmathsetmacro{\tjump}{0.535762818744}
\pgfmathsetmacro{\Tc}{537}

\definecolor{android_blue}{RGB}{51,181,229}
\definecolor{android_dark_blue}{RGB}{0,153,204}
\definecolor{android_pink}{RGB}{170,102,204}
\definecolor{android_purple}{RGB}{156,39,176}
\definecolor{android_dark_pink}{RGB}{153,51,204}
\definecolor{android_green}{RGB}{153,204,0}
\definecolor{android_dark_green}{RGB}{102,153,0}
\definecolor{android_orange}{RGB}{255,152,0}
\definecolor{android_dark_orange}{RGB}{255,152,0}
\definecolor{android_red}{RGB}{255,68,68}
\definecolor{android_dark_red}{RGB}{204,0,0}
\definecolor{android_pink}{RGB}{156,39,176}
\definecolor{android_grey}{RGB}{158,158,158}

\pgfplotsset{grid style={dashed,grey,opacity=0.5}}
	  
\pgfplotscreateplotcyclelist{peak_temp}{
	  {color=android_green,line width=0.5pt,mark size=2pt,mark options={line width=0.75pt},line join=round},
	  {color=android_red,line width=0.5pt,mark size=2pt,mark options={line width=0.75pt},line join=round},
	  {color=android_blue,line width=0.5pt,mark size=2pt,mark options={line width=0.75pt},line join=round},
	  {color=android_pink,line width=0.5pt,mark size=2pt,mark options={line width=0.75pt},line join=round},
	  {color=android_orange,line width=0.5pt,mark size=2pt,mark options={line width=0.75pt},line join=round},
	  {color=black,line width=0.5pt,mark size=2pt,mark options={line width=0.75pt},line join=round}}

\begin{document}

\title{AC noise reduction based on exchange coupled grains for heat-assisted-magnetic recording: The effect of an FeRh interlayer} 

\author{Christoph Vogler}
\email{christoph.vogler@tuwien.ac.at}
\affiliation{Faculty of Physics, University of Vienna, Boltzmanngasse 5, 1090 Vienna, Austria}
\affiliation{Institute of Analysis and Scientific Computing, TU Wien, Wiedner Hauptstrasse 8-10, 1040 Vienna,\,Austria}

\author{Claas Abert}
\author{Florian Bruckner}
\author{Dieter Suess}
\affiliation{Christian Doppler Laboratory for Advanced Magnetic Sensing and Materials, Faculty of Physics, University of Vienna, Boltzmanngasse 5, 1090 Vienna, Austria}

\begin{abstract}
High storage density and high data rate are two of the most desired properties of modern hard disk drives. Heat-assisted magnetic recording (HAMR) is believed to achieve both. Recording media, consisting of exchange coupled grains with a high and a low $T_{\mathrm{C}}$ part, were shown to have low DC noise, but increased AC noise, compared to hard magnetic single phase grains, like FePt~\cite{vogler_areal_2016}. In this work we extensively investigate the influence of an FeRh interlayer on the magnetic noise in exchange coupled grains. We find an optimal grain design that reduces the jitter in down-track direction by up to 30\,\% and in off-track direction by up to 50\,\%, depending on the head velocity, compared to the same structures without FeRh. Further, the mechanisms causing this jitter reduction are demonstrated. Additionally, we show that for ultrashort heat pulses and low write temperatures the switching time distribution of the analyzed grain structure is reduced by a factor of four, compared to the same structure without FeRh layer. This feature could be interesting for HAMR with a pulsed laser spot and could resume the discussion about this HAMR technique.
\end{abstract}

\maketitle 

\pgfplotsset{colormap/RdBu-9}

\section{Introduction}
\label{sec:intro}

The areal storage density (AD) of conventional perpendicular magnetic recording (PMR) has shown a tremendous increase over decades. Almost twenty years ago, when the annual AD growth was about 100\,\%, the limit of 1\,Tb/in$^2$ was predicted for PMR~\cite{wood_feasibility_2000}. Indeed, in recent years the growth rate has significantly slowed down and seems to reach saturation. Current AD of PMR media are already in the vicinity of 1\,Tb/in$^2$. New approaches like two-dimensional magnetic recording and/or shingled recording were proposed to extend this limit to about 1.5\,Tb/in$^2$~\cite{wood_feasibility_2009,weller_l10_2013}. 

Nevertheless, to reach even higher AD a new recording concept is needed. Heat-assisted magnetic recording (HAMR) is expected to be this key technology. Although the basic principle of HAMR was proposed nearly 60 years ago~\cite{mayer_curiepoint_1958} only recent advances in the development of near field transducers allow to fabricate HAMR heads for high density recording. HAMR uses the decrease of the coercivity of a ferromagnetic material with temperature. Hence, magnetic grains with a high magnetic anisotropy can be switched, which would not be possible at room temperature. As a consequence, recording grains can be further scaled down, while maintaining their long-term stability. The loss of long-term stability for small structures, known as superparamagnetic limit, is the bottleneck of PMR. Even more importantly, due to the heat assist the higher effective field gradient of a HAMR head allows to narrow bit transitions, ultimately yielding smaller bits.

First working devices with an AD of 1\,Tb/in$^2$~\cite{wu_hamr_2013} in 2013 and 1.4\,Tb/in$^2$~\cite{ju_high_2015} in 2015 proved the potential of HAMR. Future hard disk drives based on granular media are hoped to reach AD of up to 4\,Tb/in$^2$~\cite{wang_hamr_2013,weller_hamr_2014,weller_review_2016}. In combination with bit-patterned media theoretical investigations even predict AD beyond 10\,Tb/in$^2$~\cite{vogler_heat-assisted_2016}. 

The downside of the latter study is the low head velocity of 7.5\,m/s for which the high AD was obtained. For higher data rates, DC noise arises at high temperatures. The reason is the combination of a high thermal gradient of the heat spot and a high $\mathrm{d}H_{\mathrm{c}}/\mathrm{d}T$ gradient near the Curie temperature of the used high anisotropy islands (like FePt). Both result in a short recording time window~\cite{zhu2013understanding,zhu2015medium}, in which the external field can efficiently reverse the magnetization of the islands~\cite{vogler_basic_2016}. Hence, the switching is not reliable at high data rates. One possible solution to increase the head velocity, without getting problems with DC noise is the use of exchange coupled grains with a high and a low $T_{\mathrm{C}}$ part, lowering the $\mathrm{d}H_{\mathrm{c}}/\mathrm{d}T$ gradient. However, with such grains AC noise increases, which significantly decreases the maximum AD~\cite{vogler_areal_2016}. 

In this work we present how an FeRh layer in the middle of a high/low $T_{\mathrm{C}}$ grain (Py/FePt for example) can reduce AC noise, without suffering from DC noise, even for high data rates. Further, we investigate which requirements the FeRh interlayer must fulfill to gain the most from the proposed structure in comparison to ordinary high/low $T_{\mathrm{C}}$ grains.

\section{Methods}
\label{sec:methods}

To correctly model the magnetization dynamics during the HAMR process we use a coarse grained model based on the Landau-Lifshitz-Bloch (LLB) equation~\cite{garanin_thermal_2004,chubykalo-fesenko_dynamic_2006,atxitia_micromagnetic_2007,kazantseva_towards_2008,schieback_temperature_2009,bunce_laser-induced_2010,mendil_resolving_2014}. In detail the stochastic formulation proposed by Evans~et~at.~\cite{evans_stochastic_2012} per:
\begin{eqnarray}
\label{eq:LLB}
  \frac{d \boldsymbol{m}}{dt}= &-&\mu_0{\gamma'}\left( \boldsymbol{m}\times \boldsymbol{H}_{\mathrm{eff}}\right) \nonumber \\
  &-&\frac{\alpha_\perp\mu_0 {\gamma'}}{m^2} \left \{ \boldsymbol{m}\times \left [ \boldsymbol{m}\times \left (\boldsymbol{H}_{\mathrm{eff}}+\boldsymbol{\xi}_{\perp}  \right ) \right ] \right \}\nonumber \\
  &+&\frac{\alpha_\parallel  \mu_0{\gamma'}}{m^2}\boldsymbol{m}\left (\boldsymbol{m}\cdot\boldsymbol{H}_{\mathrm{eff}}  \right )+\boldsymbol{\xi}_{\parallel}.
\end{eqnarray}
is solved. In this equation $\boldsymbol{m}$ is the reduced magnetization $\boldsymbol{M}/M_0$, with the saturation magnetization at zero temperature $M_0$, $\gamma'$ is the reduced electron gyromagnetic ratio ($\gamma'=|\gamma_{\mathrm{e}}|/(1+\lambda^2)$ with $|\gamma_{\mathrm{e}}|=1.76086\cdot10^{11}$\,(Ts)$^{-1}$), $\mu_0$ is the vacuum permeability and $\alpha_\parallel$ and $\alpha_\perp$ are dimensionless temperature dependent longitudinal and transverse damping parameters defined as:
\begin{equation}
\label{eq:alpha}
 \alpha_\perp=\begin{cases}\lambda\left( 1-\frac{T}{T_{\mathrm{C}}} \right) & T<T_{\mathrm{C}}\\ \alpha_\parallel & T\geq T_{\mathrm{C}}\end{cases},\quad\alpha_\parallel=\lambda \frac{2T}{3T_{\mathrm{C}}}.
\end{equation}
$\lambda$ denotes the coupling of the spin to the heat bath on an atomistic level. The effective magnetic field $\boldsymbol{H}_{\mathrm{eff}}$ comprises external field, anisotropy field and exchange fields. Magnetostatic interactions are not directly included. The stochastic influence of temperature is represented by the longitudinal and perpendicular stochastic fields $\boldsymbol{\xi}_{\parallel}$ and $\boldsymbol{\xi}_{\perp}$, respectively. Both are vectors with Gaussian random numbers with zero mean and a variance per:
\begin{equation}
  \left \langle \xi_{\eta,i}(t,\boldsymbol{r})\xi_{\eta,j}({t}',\boldsymbol{r}') \right \rangle = 2D_\eta \delta_{ij}\delta(\boldsymbol{r}-\boldsymbol{r}')\delta(t-{t}'),
\end{equation}
with the diffusion constants $D_\eta$ ($\eta$ is a placeholder for $\parallel$ and $\perp$), which are derived from the fluctuation-dissipation theorem per:
\begin{eqnarray}
\label{eq:variance}
  D_\perp&=&\frac{\left (\alpha_\perp-\alpha_\parallel  \right )k_{\mathrm{B}} T}{ \gamma' \mu^2_0 M_0 V \alpha^2_\perp}\nonumber\\
  D_\parallel&=&\frac{\alpha_\parallel \gamma' k_{\mathrm{B}} T}{M_0 V}.
\end{eqnarray}
Here, $T$ is the temperature, $k_{\mathrm{B}}$ is the Boltzmann constant and $V$ is the discretization volume. The special feature of the coarse grained LLB model is that each material layer in a magnetic grain is described with a single magnetization vector. Although this coarse grained model is computationally cheap, it can correctly reproduce the magnetization dynamics of the same system with an atomistic discretization. Please refer to Ref.~\cite{volger_llb} for more details.

\subsection{HAMR setup}
\label{sec:HAMRsetup}
We analyze the switching probabilities of continuous HAMR in the following. This means, a constant laser pulse with a write temperature of $T_{\mathrm{write}}=715$\,K in the spot center and a full width at half maximum FWHM of 20\,nm is assumed to move over a recording medium. The heat pulse has a Gaussian shape in space and thus also in time. Further, we assume the inductive write head to produce a spatially uniform field with a strength of $ \mu_0 H_{\mathrm{ext}}=0.8$\,T. We choose a typical write frequency of 1\,GHz. The easy axis of the grains and the external field enclose an angle of 6\,$^\circ$.
With this setup we investigate the switching behavior of isolated grains with an initial magnetization state that points in the negative $z$ direction, referred to as magnetization down in the following. Depending on the off-track position of a grain in the medium the maximum temperature of the arriving heat pulse differs per:
\begin{equation}
\label{eq:peak_temp_CLSR}
 T_{\mathrm{max}}(y)=\left ( T_{\mathrm{write}}-T_{\mathrm{min}} \right )e^{-\frac{y^2}{2\sigma^2}}+T_{\mathrm{min}},
\end{equation}
with $y$ being the off-track position, $T_{\mathrm{min}}=270$\,K and 
\begin{equation}
 \sigma=\frac{\mathrm{FWHM}}{\sqrt{8\ln(2)}}.
\end{equation}
The maximum temperature of the heat pulse is independent from the down-track direction but the temporal shift between the maximum of the heat pulse and the center of the write pulse varies. 
\begin{figure}
\includegraphics{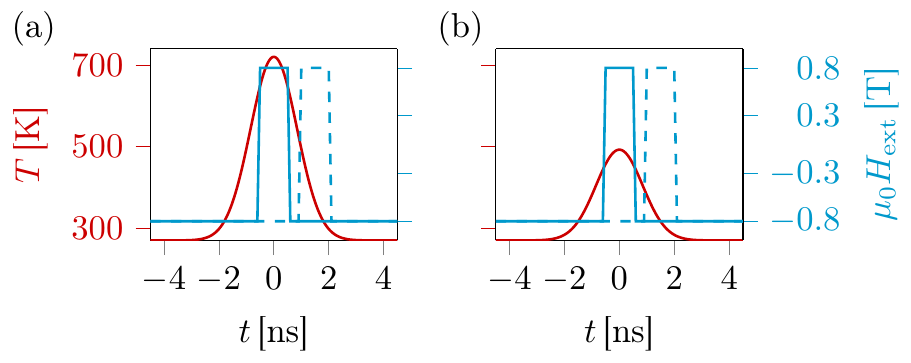}
    \caption{\small (color online) Illustration of the temporal evolution of the applied heat pulse and the external magnetic field for a write pattern of 0001000 for different grain positions. A head velocity of 10\,m/s is assumed. (a) Grain at the center of the track (off-track position of $y=0$\,nm) and a down-track position of $x=0$\,nm (solid blue) and $x=15$\,nm (dashed blue), respectively. (b) Grain at the same down-track positions as in (a) but an off-track position of $y=10$\,nm, yielding a lower maximum temperature of the heat pulse.}
  \label{fig:setup}
\end{figure}
In order to illustrate the situation for different bit positions, Fig.~\ref{fig:setup} displays the temporal evolution of the heat pulse and the external field for a write pattern of 0001000. 

\subsection{recording grains}
\label{sec:grains}
\begin{table}[h!]
  \centering
  \vspace{0.5cm}
  \begin{tabular}{c c c c}
    \toprule
    \toprule
      & HM  & SM1 & SM2 \\
    \midrule
    $K_1$\,[MJ/m$^3$] & $6.6$ & 0.0 & 0.0\\
    $\mu_0 M_{\mathrm{S}}$\,[T] & 1.43 & 1.43 & 1.00 \\
    $\lambda$ & 0.1 & 1.0 & 1.0\\
    $T_{\mathrm{C}}$\,[K] & 537 & 822 & 849 \\
    \bottomrule
    \bottomrule
  \end{tabular}
  \caption{\small Zero temperature magnetic properties of the used soft magnetic (SM) and hard magnetic (HM) materials. $K_1$ is the anisotropy constant, $M_{\mathrm{S}}$ the saturation magnetization, $\lambda$ the spin-bath coupling strength and $T_{\mathrm{C}}$ is the Curie temperature.}
  \label{tab:mat_match}
\end{table}
The grains in the investigated media have a diameter of 5\,nm and a thickness of 10\,nm. We compare three grain types:
\begin{itemize}
 \item single phase grains with the material properties HM of Table~\ref{tab:mat_match}. 
 \item exchange coupled SM/HM grains consisting of 50\,\% SM (SM1 or SM2) and 50\,\% HM material. 
 \item SM/FeRh/HM grains with the same composition as the latter but with an FeRh interlayer in between SM and HM layers.
\end{itemize}
The HM material of Table~\ref{tab:mat_match} is very similar to pure FePt, except for the reduced Curie temperature. A decrease of $T_{\mathrm{C}}$, while maintaining high anisotropy, can be obtained by adding a few percent of Cu or Ni as demonstrated in Refs.~\cite{thiele_temperature_2002,wang_re-evaluation_2011,gilbert_tuning_2013}. The SM layers have properties similar to those of Permalloy (Py) with enhanced damping constant. Although this assumption seems risky, Bailey~et~\textit{al.}~\cite{bailey_control_2001} have shown how to significantly increase the damping constant in Py while maintaining its soft magnetic character by doping with low concentrations of Tb.

At room temperature FeRh is antiferromagnetic. Above a critical temperature $T_{\mathrm{jump}}$ it becomes ferromagnetic. The key idea of the proposed structure is to use a thin FeRh layer between the SM and the HM layer, which controls the exchange coupling strength between the latter. In our model we do not compute the detailed magnetization dynamics of the FeRh interlayer, we only consider the effected modulation of the exchange coupling, as demonstrated in Fig.~\ref{fig:FeRh_sketch}b. In the ferromagnetic phase, well above  $T_{\mathrm{jump}}$, we assume full coupling and towards room temperature the coupling drastically decreases.

In contrast to FeRh/HM grains without the soft top layer, in which FeRh is used to produce an exchange spring effect above $T_{\mathrm{jump}}$~\cite{thiele_ferh/fept_2003}, FeRh just tunes the exchange coupling in the proposed SM/FeRh/HM structure. FeRh itself is not required to lower the coercivity of the composite grain. For the read-back process the additional magnetic moment of the SM layer is a large advantage to achieve high signal-to-noise ratios even for small grain dimensions. This is also in contrast to the FePt/FeRh/FeCo trilayer structure of Refs.~\cite{zhu_binary_2008,zhou_concept_2012} with an in-plane FeCo layer, whose magnetic moment cannot be used during read-back. 

\subsection{SM layer stabilization}
\label{sec:layer_stabilization}
In the proposed SM/FeRh/HM structure it must be ensured that the magnetization of the SM layer is orientated in the correct direction (parallel to that of the HM layer) during read-back. Before discussing the recording performance of the introduced grains we want to analyze the requirements under which the proposed SM/FeRh/HM structures show this parallel alignment, and thus a high read-back signal, due to the high moment of the SM part, is ensured. 

In equilibrium the exchange field of the HM layer, acting on the SM layer, must be larger than the sum of the strayfields of all neighboring grains and the strayfield of the bottom HM layer. This means that the exchange coupling between SM layer and HM layer must not completely vanish at room temperature. As a consequence it is required that the FeRh interlayer must not become a perfectly compensated antiferromagnet. To estimate the required room temperature coupling strength we calculate the strayfield of all neighboring grains and the bottom HM layer for a worst case scenario. In this scenario all grains in an area of 50\,nm times 50\,nm of a granular medium are assumed to be magnetized in up direction, trying to reverse the SM layer. The surface of the medium consists of 83\,\% grains and 17\,\% grain boundary which corresponds to an average boundary thickness of 0.5\,nm for grains with an average diameter of 5\,nm. We generate 1000 different realizations of the described sector via Voronoi tessellation and compute the average strayfield in the top part of the center grain with \textit{magnum.fe}~\cite{abert_magnumfe_2013}. In detail, a hybrid finite element - boundary element method (FEM/BEM) based on the Fredkin/Koehler approach~\cite{fredkin_hybrid_1990} is used for the computation of the strayfield. 
\begin{figure}
   \includegraphics{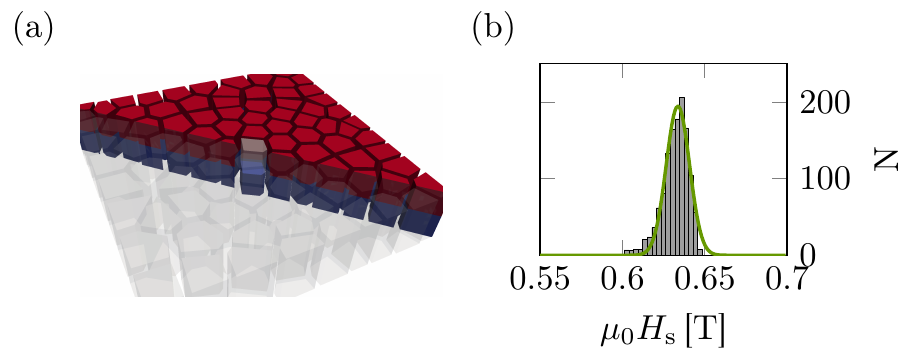}
      \caption{\small (color online) (a) Slice through a medium with an area of 50x50\,nm$^2$, consisting of SM/HM grains with an average diameter of 5\,nm and a thickness of 10\,nm. The colors are used to differ between the HM and SM parts of the grains. The strayfield in the upper highlighted SM part of the center grain is evaluated. (b) Strayfield distribution in the upper highlighted SM1 part of the center grain, produced by all neighboring grains and the lower HM part of the center grain. The distribution is computed from 1000 different media realizations.}
      \label{fig:medium}  
\end{figure}
Figure~\ref{fig:medium}a demonstrates a slice of one realization of the medium. The colors are used to differ between the HM and SM parts of the grains. A histogram of the strayfield in the top SM1 part of the center grain, which is highlighted in Fig.~\ref{fig:medium}a, is illustrated in Fig.~\ref{fig:medium}b. The resulting maximum and average demagnetizing fields of both HM/SM structures are given in Table~\ref{tab:demag}.
\begin{table}[h!]
  \centering
  \vspace{0.5cm}
  \begin{tabular}{c c c c}
    \toprule
    \toprule
    grain & $A_{\mathrm{iex}}(0)$ & $\mu_0 H_{\mathrm{s,av}}$ & $\mu_0 H_{\mathrm{s,max}}$ \\
    \midrule
    SM1/HM & 25.75\,pJ/m & 634\,mT & 648\,mT \\
    SM2/HM & 25.75\,pJ/m & 604\,mT & 614\,mT \\
    \bottomrule
    \bottomrule
  \end{tabular}
  \caption{\small Average and maximum strayfield in the soft magnetic part of the center grain as illustrated in Fig.~\ref{fig:medium}. $A_{\mathrm{iex}}(0)$ denotes the full exchange coupling between soft and hard magnetic layer at zero temperature.}
  \label{tab:demag}
\end{table}
According to Ref.~\cite{volger_llb} the intergrain exchange field of the HM part acting on the SM part can be computed per:
\begin{equation}
\label{eq:Hiex}
 H_\mathrm{iex,SM}(T)=\frac{2 A_{\mathrm{iex}}(T)}{at\mu_0 M_{\mathrm{S,SM}}(T)},
\end{equation}
where $a$ is the atomistic lattice constant, $t$ the layer thickness and $M_{\mathrm{S,SM}}(T)$ the saturation magnetization of the SM layer at the temperature $T$.
\begin{figure}
   \includegraphics{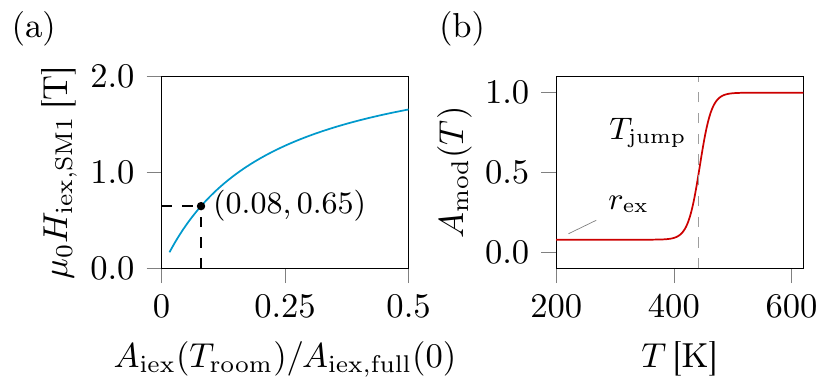}
      \caption{\small (color online) (a) Intergrain exchange field of a HM layer acting on a SM1 layer as function of the reduction of the exchange constant between the latter. The exchange constant is normalized to the full exchange at zero temperature. (b) Modulation function of the intergrain exchange of an FeRh interlayer in a SM1/FeRh2/HM structure (see Table~\ref{tab:FeRH_param}). $r_{\mathrm{ex}}$ denotes the remaining fraction of exchange at low temperatures due to a remaining net magnetic moment of FeRh and $T_{\mathrm{jump}}$ denotes the critical jump temperature of the first order phase transition in FeRh.}
    \label{fig:FeRh_sketch}  
\end{figure}
The obtained reduction of the intergrain exchange field acting on the SM1 layer with decreasing coupling strength is illustrated in Fig.~\ref{fig:FeRh_sketch}a. To ensure that the magnetization of the SM layer points in the same direction as the HM layer magnetization we must claim that $H_{\mathrm{iex,SM}} \ge H_{\mathrm{s,max}}$ is valid at room temperature. This holds if the remaining exchange coupling in the SM/FeRh/HM structure at room temperature is larger than $r_{\mathrm{ex}}A_{\mathrm{iex,full}}(0)$. $r_{\mathrm{ex}}$ depends on the composition of the grains.

The above considerations suggest that the used FeRh interlayer must meet some requirements, like a remaining ferromagnetic moment at room temperature. It was demonstrated that the remaining room temperature ferromagnetic moment of FeRh depends on the annealing temperature~\cite{ohtani_antiferroferromagnetic_1998,cao_magnetization_2008}, on the composition~\cite{kande_origin_2010}, on the film thickness and also on the extent of atomic ordering~\cite{kande_enhanced_2011}. Hence, for small films it seems to be adjustable. As shown in the following also the critical temperature of the first order phase transition $T_{\mathrm{jump}}$ is an important quantity for the proposed structure. $T_{\mathrm{jump}}$ is known to be tunable in a wide temperature range~\cite{kouvel_unusual_2004,thiele_ferh/fept_2003}. To quantify the exchange modulation function of the FeRh layer we use~\cite{ostler_modeling_2017}:
\begin{equation}
\label{eq:FeRh_exchange}
 \frac{A_{\mathrm{iex}}(T)}{A_{\mathrm{iex}}(0)}=\alpha\tanh\left( \frac{T-T_{\mathrm{jump}}}{\Delta T}\right)+\beta.
\end{equation}
Here, $\Delta T$ denotes the width of the transition, $\alpha$ is a scaling parameter and $\beta$ is an offset parameter. According to measurements of FeRh thin films in Refs.~\cite{kande_enhanced_2011,ostler_modeling_2017} we use a typical value of $\Delta T=20$\,K. The scaling and offset parameters are computed from the requirement of the remaining exchange coupling fraction $r_{\mathrm{ex}}$ well below $T_{\mathrm{jump}}$ per:
\begin{eqnarray}
\label{eq:FeRh_exchange_a}
 \alpha&=&\frac{1-r_{\mathrm{ex}}}{1-\tanh \left( -\frac{T_{\mathrm{jump}}}{\Delta T} \right)},\nonumber \\
 \beta&=&r_{\mathrm{ex}}-\alpha \tanh \left( -\frac{T_{\mathrm{jump}}}{\Delta T} \right).
\end{eqnarray}
The parameters for all examined SM/FeRh/HM structures are given in Table~\ref{tab:FeRH_param}.
\begin{table}[h!]
  \centering
  \vspace{0.5cm}
  \begin{tabular}{l c c c c c}
    \toprule
    \toprule
    grain & $T_{\mathrm{jump}}$\,[K] & $\Delta T$\,[K] & $r_{\mathrm{ex}}$\,[\%] & $\alpha$ & $\beta$ \\
    \midrule
    SM1/FeRh1/HM & 442 & 20 & 0.0 & 0.50 & 0.50 \\
    SM1/FeRh2/HM & 444 & 20 & 8.0 & 0.46 & 0.54 \\
    SM2/FeRh3/HM & 466 & 20 & 6.7 & 0.47 & 0.53 \\
    \bottomrule
    \bottomrule
  \end{tabular}
  \caption{\small Parameters, according to Eqs.~\ref{eq:FeRh_exchange} and \ref{eq:FeRh_exchange_a}, defining the intergrain exchange modulation functions of FeRh in three different grain structures.}
  \label{tab:FeRH_param}
\end{table}

Note, that there are many requirements on the FeRh layer in the proposed structure. It is not clear if all of them can be fulfilled. However, in this work we aim to investigate the potential benefits of SM/FeRh/HM grains in HAMR. Hence, we assume the desired properties. Further, the presented structures can serve as design guidelines for experimental research on FeRh.

\section{Results}
\label{sec:Results}
\subsection{footprints}
\label{sec:footprints}
\begin{figure*}
\begin{adjustwidth}{-1cm}{-1cm}
\includegraphics{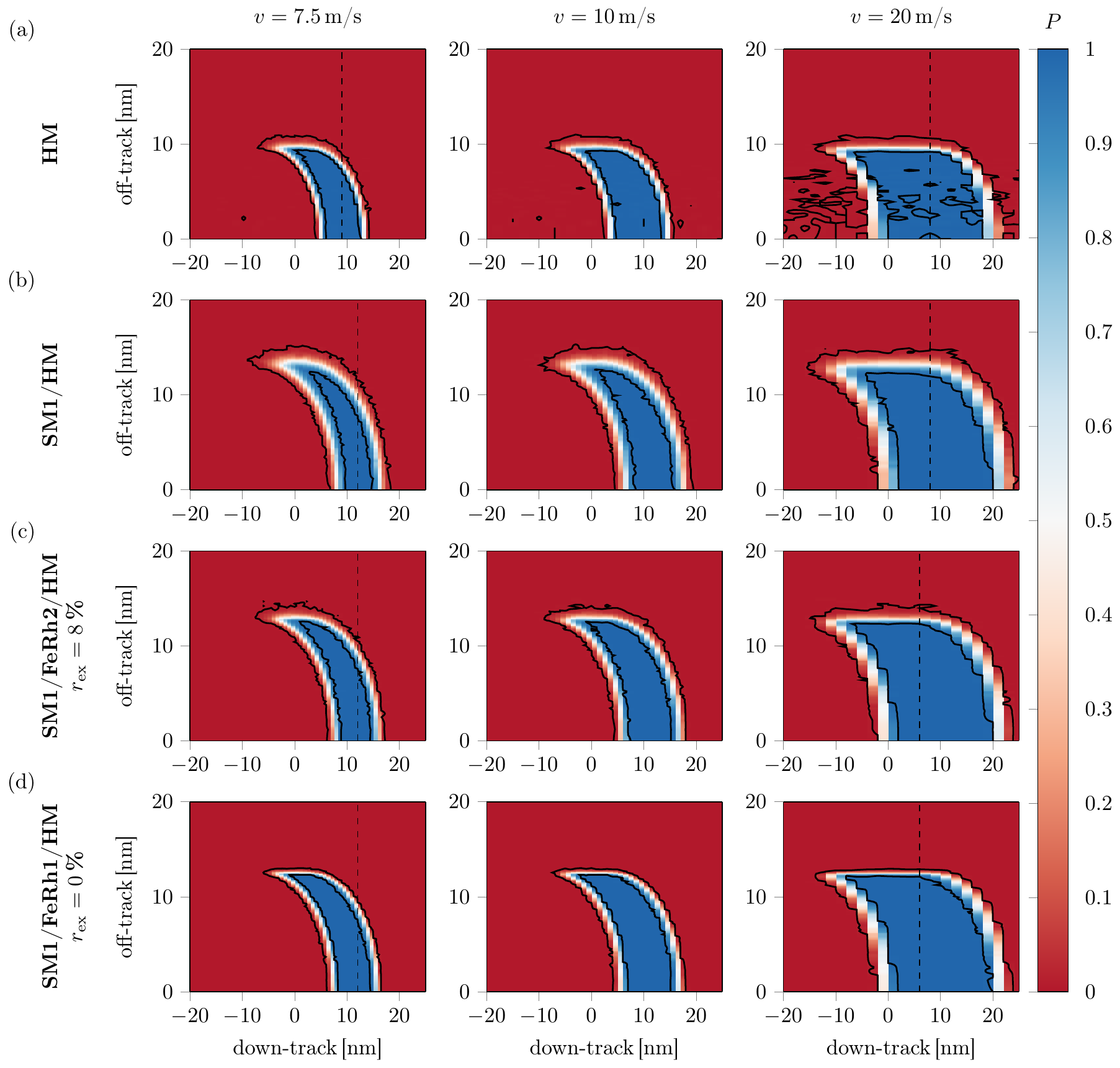}
  \caption{\small (color online) Footprint of a HAMR head for three different velocities, writing on various grain types. The color code illustrates the switching probability $P$ of a grain located at the corresponding position. $P$ is computed from 128 switching trajectories, which try to reverse the magnetization from an initial down to an up direction. A FWHM of 20\,nm and an external field strength of 0.8\,T is assumed for all phase points and grain types. The contour lines separate the areas with 0\,\% and 100\,\% switching probability. Along the vertical dashed lines the off-track jitter displayed in Table~\ref{tab:jitter} is computed.}
  \label{fig:footprints}
\end{adjustwidth}
\end{figure*}
By means of the coarse grained LLB model introduced in Sec.~\ref{sec:methods} and the HAMR setup of Sec.~\ref{sec:HAMRsetup} we simulate footprints of the HAMR head on recording media consisting of the grain types described in Sec.~\ref{sec:grains}. Three different head velocities are examined as displayed in Fig.~\ref{fig:footprints}. It is assumed that a grain is located at each phase point of Fig.~\ref{fig:footprints} and 128 write trials are computed for each position. The switching probability $P$ is finally obtained from the fraction of successfully reversed grains. In the case of bit-patterned media the presented phase diagrams directly determine the switching probability of bits and in the case of granular media they can be interpreted as average footprints of the head. Hence, for both kind of media conclusions can be drawn. 

Figure~\ref{fig:footprints}a shows narrow transitions for a head velocity of 7.5\,m/s and pure HM grains. For bit-patterned media consisting of the same HM grains Ref.~\cite{vogler_heat-assisted_2016} demonstrated a maximum areal storage density of about 13\,Tb/in$^2$ if a shingled write schema is used. Although the transitions remain narrow at higher head velocities DC noise becomes a serious problem. The reason is the combination of a high thermal gradient and a high head velocity, which both reduce the recording time window of the write process. According to Ref.~\cite{vogler_basic_2016} the effective recording time window is defined per:
\begin{equation}
 \label{eq:ERTW}
 \text{ERTW}_\uparrow=\left[t(T_{\mathrm{C}}),t(T_{\mathrm{f}})\right] \cap \left [ t_{\uparrow,\mathrm{start}},t_{\uparrow,\mathrm{final}}\right],
\end{equation}
where $T_{\mathrm{C}}$ is the Curie temperature and $T_{\mathrm{f}}$ is the freezing temperature. At $T_{\mathrm{f}}$ the coercive field of a grain becomes lower than the available write field. Hence, Eq.~\ref{eq:ERTW} depicts the intersection of the time interval during which the external field points in write direction $[ t_{\uparrow,\mathrm{start}},t_{\uparrow,\mathrm{final}}]$ and the time interval $[t(T_{\mathrm{C}}),t(T_{\mathrm{f}})]$ during which the coercive field of the grains is lower than the write field. Only during the ERTW the external field can efficiently act on the magnetization of the particles. If the recording time window is smaller than a threshold value the switching process is not reliable, yielding DC noise (see Ref.~\cite{vogler_basic_2016}), in the case of the pure HM grain and $v=20$\,m/s. 

The exchange spring SM/HM structure significantly reduces DC noise even for higher head velocities, at the expense of a much broader transition area due to the increased AC noise, as Fig.~\ref{fig:footprints}b points out. The reason is the lower $\mathrm{d}H_{\mathrm{c}}/\mathrm{d}T$ gradient near the Curie temperature, which enlarges the recording time window and thus reduces DC noise but increases AC noise. This means, in principle higher data rates are possible but the distance between neighboring bits in bit-patterned media or the transition jitter in granular media increases, resulting in lower maximum areal densities (see Ref~\cite{vogler_areal_2016}). 
\begin{figure}
   \includegraphics{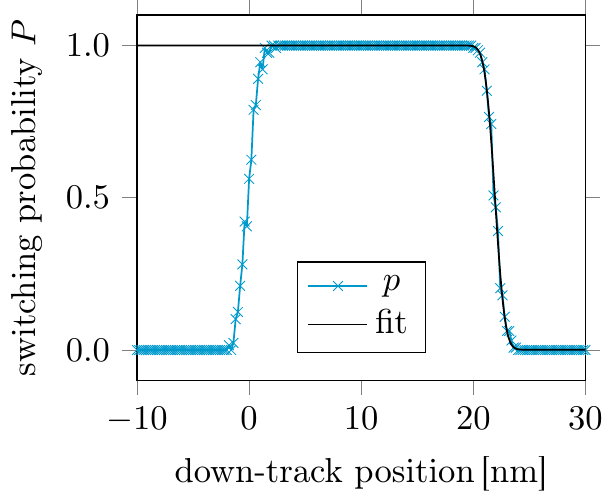}
      \caption{\small (color online) Switching probability of a SM1/HM grain for a head velocity of 20\,m/s, along the center track taken from Fig.~\ref{fig:footprints}b. The transition is fitted with the cumulative distribution function of the normal distribution to extract a measure for the transition width.}
    \label{fig:down_track_fit}  
\end{figure}

To quantify the transition width in down-track direction the switching probabilities along the center track ($y=0$ in Fig.~\ref{fig:footprints}) are fitted with the cumulative distribution function of the normal distribution as demonstrated in Fig.~\ref{fig:down_track_fit}. The obtained standard deviation $\sigma$ is a measure for the transition jitter. The same fitting procedure along the marked dashed lines in Fig.~\ref{fig:footprints} is performed to extract the transition width in off-track direction. Table~\ref{tab:jitter} summarizes the jitter values for all investigated grains and for head velocities of 7.5\,m/s and 20\,m/s, respectively.
\begin{table*}
  \centering
  \vspace{0.5cm}
  \begin{tabular}{l c | c c c c | c c c c}
    \toprule
    \toprule
    \multirow{2}{*}{grain} & \multirow{2}{*}{$r_{\mathrm{ex}}$\,[\%]} & \multicolumn{4}{c}{$v=7.5$\,m/s} & \multicolumn{4}{|c}{$v=20$\,m/s} \\
     &  & $\sigma_{\mathrm{down}}$\,[nm] &  $\sigma_{\mathrm{off}}$\,[nm] & $\sigma_{\mathrm{down,red}}$\,[\%] & $\sigma_{\mathrm{off,red}}$\,[\%] & $\sigma_{\mathrm{down}}$\,[nm] &  $\sigma_{\mathrm{off}}$\,[nm] & $\sigma_{\mathrm{down,red}}$\,[\%] & $\sigma_{\mathrm{off,red}}$\,[\%]\\
    \midrule
    HM & - & 0.20 & 0.27 & - & - & 0.39 & 0.23 & - & - \\
    SM1/HM & - & 0.55 & 0.74 & - & - & 0.70 & 0.49 & - & -\\
    SM2/HM & - & 0.62 & 0.83 & - & - & 0.70 & 0.51 & - & -\\
    \midrule
    SM1/FeRh1/HM & 0.0 & 0.33 & 0.33 & -39.19 & -55.07 & 0.52 & 0.09 & -25.86 & -81.68\\
    SM1/FeRh2/HM & 8.0 & 0.41 & 0.51 & -24.23 & -30.83 & 0.56 & 0.25 & -19.09 & -49.26\\
    SM2/FeRh3/HM & 6.7 & 0.45 & 0.54 & -27.13 & -35.01 & 0.50 & 0.27 & -27.88 & -47.83 \\
    \bottomrule
    \bottomrule
  \end{tabular}
  \caption{\small Transition jitter in down-track and off-track direction for various grain types and head velocities. The jitter reductions of the SM/FeRh/HM structures refer to the corresponding ordinary SM/HM grains.}
  \label{tab:jitter}
\end{table*}

As expected, the transition jitter of the SM/HM structures is significantly higher than in the pure HM grains in both directions. The proposed SM1/FeRh2/HM structure with a remaining ferromagnetic moment in the FeRh interlayer with $r_{\mathrm{ex}}=8$\,\% (for definition of $r_{\mathrm{ex}}$ see Eqs.~\ref{eq:FeRh_exchange} and \ref{eq:FeRh_exchange_a} as well as Fig.~\ref{fig:FeRh_sketch}b) shows a reduction of the down-track jitter of about 20-25\,\% for both examined head velocities (see Table~\ref{tab:jitter}) compared to ordinary SM1/HM grains. As Fig.~\ref{fig:footprints}c points out no DC noise appears independent from the head velocity. The off-track jitter gain is even higher with about 30\,\% in the case of $v=7.5$\,m/s and 50\,\% for $v=20$\,m/s. Very similar values are obtained for SM2/FeRh3/HM grains with $r_{\mathrm{ex}}=6.7$\,\%. SM2/HM grains show slightly higher jitter values compared to SM1/HM structures with a higher saturation magnetization in the soft magnetic part, but the jitter reduction is almost the same for both analyzed grain types, if a suitable FeRh interlayer is used. Table~\ref{tab:jitter} demonstrates that the jitter reduction can be notably increased if the FeRh interlayer is a perfect antiferromagnet at room temperature ($r_{\mathrm{ex}}=0$\,\%). The reason will become clear in Sec.~\ref{sec:origin}. This structure is investigated to point out the maximum possible gain of the interlayer. It could not be used in a real device due to the decoupled soft magnetic layer at room temperature, which would deteriorate the read-back signal.

In a nutshell, the presented footprints of a HAMR head and the corresponding jitter calculations show that a medium consisting of well-designed SM/FeRh/HM grains can significantly reduce the transition jitter in both, down-track and off-track direction, while maintaining a low DC error. At high temperatures the structure behaves like an ordinary SM/HM grain, because well above $T_{\mathrm{jump}}$ the exchange coupling between SM and HM layer has full strength. Hence, DC noise does not arise even at high head velocities. Although, the transitions in pure HM grains are narrower, SM/FeRh/HM grains could be a good compromise for intermediate to high AD, if the requirements on the FeRh interlayer can be met.

\subsection{origin of jitter reduction}
\label{sec:origin}
To understand the reason for the reduction of the transition jitter presented in Sec.~\ref{sec:footprints} the dependence of the freezing temperature $T_{\mathrm{f}}$ on the intergrain exchange coupling $A_{\mathrm{iex}}$ between SM and HM layer must be analyzed. For this purpose we compute the temperature dependent coercivity of the SM/HM grains for various values of $A_{\mathrm{iex}}$ by means of Eq.~\ref{eq:LLB}.
\begin{figure} 
   \includegraphics{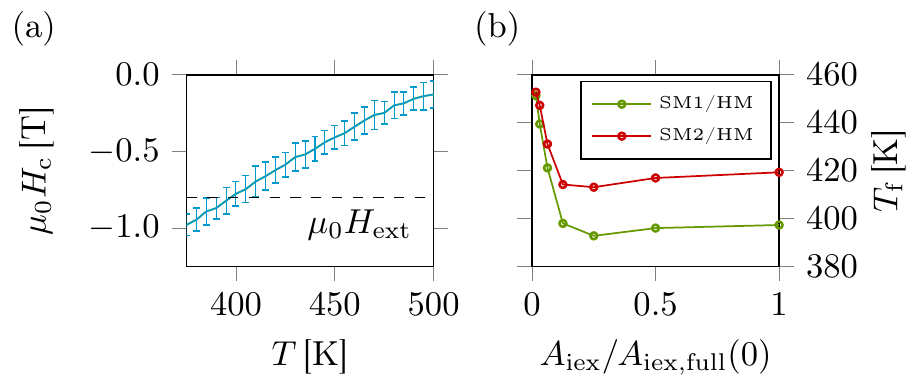}
      \caption{\small (color online) (a) Coercivity as a function of temperature for a SM1/HM grain. Each data point (value and standard deviation) is obtained from 100 hysteresis loops at a constant temperature and with a sweep rate of 100\,mT/ns. (b) Freezing temperature $T_{\mathrm{f}}$ versus the exchange coupling between the soft magnetic and the hard magnetic layer for SM1/HM and SM2/HM grains, if an external field of 0.8\,T is applied. The exchange coupling is normalized to the full interaction strength at zero temperature.}
    \label{fig:Tf_dependence}  
\end{figure}
Exemplarily, Fig.~\ref{fig:Tf_dependence}a illustrates the temperature dependent coercivity of a SM1/FeRh/HM grain with full intergrain exchange coupling (25.75\,pJ/m). To obtain $H_{\mathrm{c}}$ at each temperature repeated easy axis hysteresis loops are simulated by means of the stochastic LLB equation. Each hysteresis loop is simulated at a constant temperature and a sweep rate of 100\,mT/ns. Statistics of 100 loops yield values of the average coercivity and the corresponding standard deviation. The freezing temperature can be evaluated as the temperature at which $H_{\mathrm{c}} = H_{\mathrm{ext}}$. Figure~\ref{fig:Tf_dependence}b displays $T_{\mathrm{f}}$ for both investigated SM/HM structures and for various intergrain exchange couplings. Both grain types show the same qualitative behavior. $T_{\mathrm{f}}$ is almost constant over a wide range of exchange couplings. Only for very small couplings the freezing temperature increases. The minimum value of $T_{\mathrm{f}}$ differs based on the saturation magnetization of the SM part. This dependence of the coercivity on the intergranular exchange is well known for thin exchange spring structures~\cite{kapoor_effect_2006,suess_effect_2009,suess_superior_2016}.

The reduction of the off-track jitter in Sec.~\ref{sec:footprints} can now be understood by this increase of $T_{\mathrm{f}}$ with small exchange couplings. 
\begin{figure}
\includegraphics{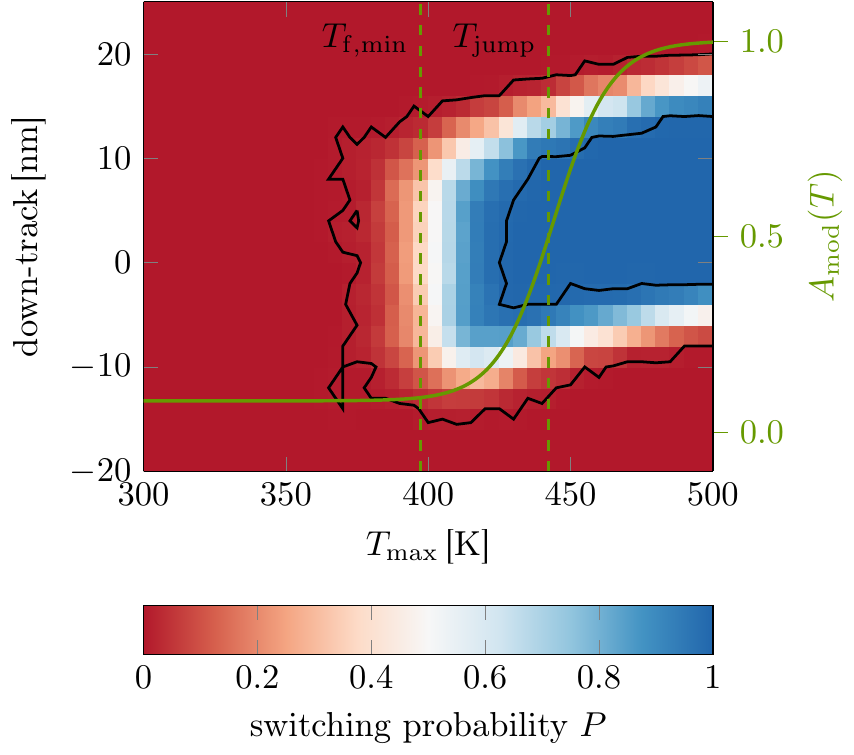}
  \caption{\small (color online) Footprint of SM1/HM grains and a head velocity of 20\,m/s taken from Fig.~\ref{fig:footprints}b. The off-track axis is replaced with the maximum temperature of the heat pulse, arising at the corresponding off-track position of the grains. The intergrain exchange modulation function of Fig.~\ref{fig:FeRh_sketch}b is plotted as an overlay.}
  \label{fig:footprint_Tf}
\end{figure}
To make this clear Fig.~\ref{fig:footprint_Tf} again shows the footprint of an HMAR head with $v=20$\,m/s writing on SM1/HM grains (see also Fig.~\ref{fig:footprints}b). In this plot the off-track direction is replaced by the maximum temperature of the heat pulse which reaches the grain (see Eq.~\ref{eq:peak_temp_CLSR}). Additionally, the modulation function of the exchange coupling of Fig.~\ref{fig:FeRh_sketch}b is displayed as an overlay. $T_{\mathrm{jump}}$ of the FeRh interlayer is slightly above the minimum freezing temperature for full coupling $T_{\mathrm{f,min}}$. As a consequence, at the off-track edge of the footprint, the exchange coupling is already small enough that the freezing temperature of the structure increases (see Fig.~\ref{fig:Tf_dependence}b). Hence, the heat-assist is too low to switch the grains and the border of the transition shifts to higher temperatures. At $T_{\mathrm{jump}}$ still 50\,\% of the full exchange coupling is available. Based on the findings of Fig.~\ref{fig:Tf_dependence}b the SM1/FeRh/HM structure still has the full exchange spring effect and the footprint does not change compared to the ordinary SM1/HM structure. This means that the width of the transition becomes smaller, which is confirmed by the off-track jitter reduction pointed out in Table~\ref{tab:jitter}. 

The same mechanism, of increasing $T_{\mathrm{f}}$ at low temperatures or weak exchange coupling, is also responsible for the jitter reduction in down-track direction. During cooling of a grain the exchange interaction decreases from full to the minimum coupling, and thus $T_{\mathrm{f}}$ increases. Based on Eq.~\ref{eq:ERTW} the effective recording time window becomes smaller as a consequence. A fast decreasing ERTW directly implies a limitation of the transition area, and thus a jitter reduction. Nevertheless, the down-track jitter will even not vanish if the phase transition of the FeRh layer becomes infinitely sharp ($\Delta T\rightarrow0$ in Eq.~\ref{eq:FeRh_exchange}). For $\Delta T=0$ the freezing temperature of a SM/FeRh/HM structure suddenly increases to that of a pure HM grain at $T_{\mathrm{jump}}$. As a consequence, the recording time window does not vanish, it reduces to that of a pure HM structure. Hence, also the down-track jitter reduces at most to that of a pure HM grain.

Note, that the jitter reduction increase strongly depends on the maximum increase of $T_{\mathrm{f}}$ at low temperatures. Hence, it is clear that SM1/FeRh1/HM grains with $r_{\mathrm{ex}}=0$\,\% show a larger jitter reduction. However, $r_{\mathrm{ex}}$ is limited, as discussed in Sec.~\ref{sec:grains}. In these considerations the strayfield of granular media was analyzed. For bit-patterned media, with larger spacings between the individual islands, the requirements on $r_{\mathrm{ex}}$ may be relaxed. For FeRh/HM without an additional soft layer~\cite{thiele_ferh/fept_2003} the mechanisms should be similar and larger jitter reductions can be expected. Nevertheless, such structures have problems during read-back due to the vanishing magnetic moment of FeRh at room temperature.

From the above considerations we see how important the proper design of the SM/FeRh/HM grains is. For both analyzed structures a clever choice of $T_{\mathrm{jump}}$ and $r_{\mathrm{ex}}$ yields very similar jitter reductions. On the other hand, there exists some tolerance for the production of the FeRh interlayer, if the soft magnetic layer is adapted.

\subsection{ultrashort pulses}
\begin{figure}
   \includegraphics{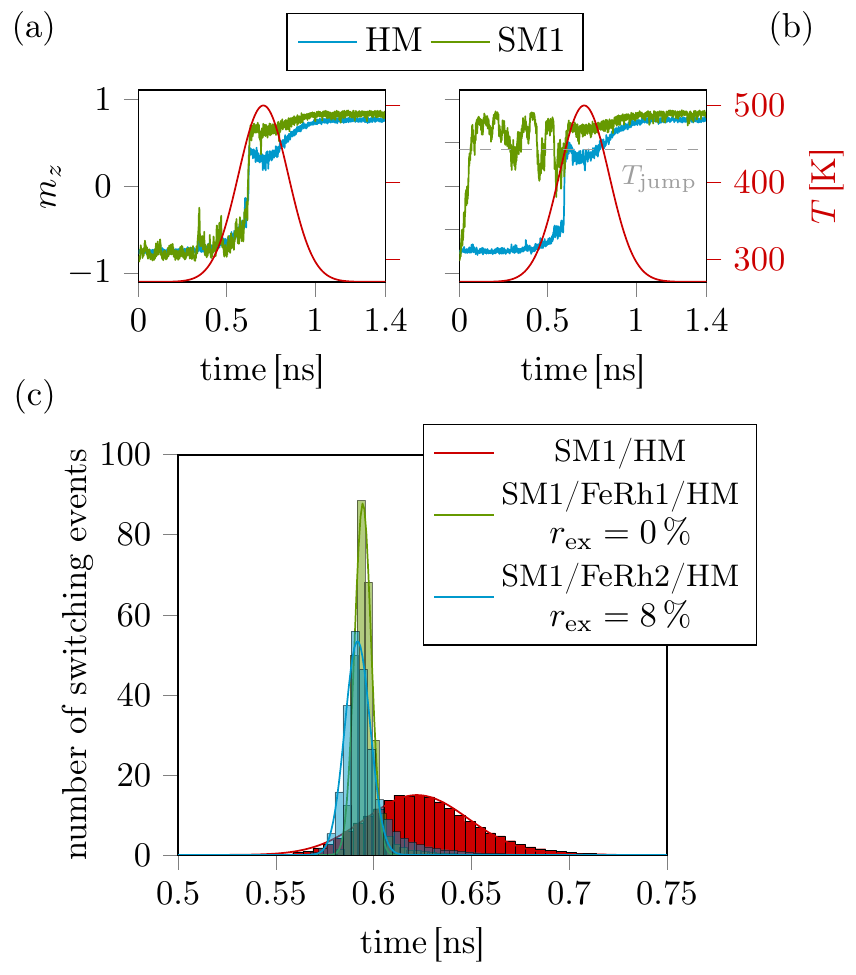}  
      \caption{\small (color online) Switching trajectories of soft and hard magnetic layer in (a) an ordinary SM1/HM (b) a SM1/FeRh2/HM structure. A constant external field with $\mu_0 H_{\mathrm{ext}}=0.8$\,T in up direction as well as a heat pulse with $\sigma_{\mathrm{pulse}}=140$\,ps and $T_{\mathrm{max}}=500$\,K is assumed. (c) Switching time distribution obtained from 50000 switching trajectories as displayed in (a) and (b) for three grain types.}
    \label{fig:distribution}  
\end{figure}
Another interesting aspect of the proposed SM/FeRh/HM structure arises if heat pulses, shorter than the field pulse duration, are used to assist the write process. In such a situation the external field can be assumed to be constant during the write process. We further assume that $T_{\mathrm{max}}$ is smaller than the Curie temperature of the HM layer. Hence, the magnetization reversal takes place during heating of the grain and not any more during cooling, because the particle never becomes paramagnetic. Under these circumstances the switching time distribution of SM/FeRh/HM grains is significantly shorter than that of the corresponding ordinary SM/HM grains, as illustrated in Fig.~\ref{fig:distribution}c. Here, the switching time of 50000 switching trajectories with $T_{\mathrm{max}}=500$\,K and and a pulse duration of $\sigma_{\mathrm{pulse}}=140$\,ps, are evaluated for three grain types. The last zero crossing of the $z$ component of the magnetization of each layer determines the corresponding switching time.
\begin{table}[h!]
  \centering
  \vspace{0.5cm}
  \begin{tabular}{l c c}
    \toprule
    \toprule
    grain & $r_{\mathrm{ex}}$\,[\%] & $\sigma_{\mathrm{sw}}$\,[ps] \\
    \midrule
    SM1/HM & - & 26.3\\
    SM1/FeRh1/HM & 0 & 4.2 \\
    SM1/FeRh2/HM & 8 & 6.5 \\
    \bottomrule
    \bottomrule
  \end{tabular}
  \caption{\small Switching time distribution of various grain types. As illustrated in Fig.~\ref{fig:distribution} the distribution for each grain type is obtained from 50000 switching trajectories with a pulse duration of $\sigma_{\mathrm{pulse}}=140$\,ps and a maximum pulse temperature of $T_{\mathrm{max}}=500$\,K. A constant external field is assumed.}
  \label{tab:switching_time_distr}
\end{table}
The standard deviations of the obtained switching time distributions are given in Table~\ref{tab:switching_time_distr}. The distribution decreases by a factor of 6 in the case of the SM1/FeRh1/HM structure with $r_{\mathrm{ex}}=0$\,\% and by a factor of 4 if $r_{\mathrm{ex}}=8$\,\% is assumed. The remaining ferromagnetic moment of the FeRh interlayer does not influence the reduction much.

The reason for the narrowing of the switching time distribution is displayed in Figs.~\ref{fig:distribution}a and b, where one randomly chosen magnetization trajectory of the HM and SM1 layer is shown for both grain types. Without FeRh interlayer the SM1/HM grain has full exchange during the whole write process. The layers are strongly coupled at all times and switch together at $T \ge T_{\mathrm{f,min}}$. In contrast, the SM1 and the HM layer decouple for the used external field with a strength of 0.8\,T at the beginning of the simulation for low temperatures. Since $H_{\mathrm{ext}}>H_{\mathrm{iex,SM1}}$ is valid the soft magnetic layer switches, whereas the hard magnetic layer does not, due to its high coercivity. Above $T_{\mathrm{jump}}$ the exchange spring effect reaches full strength. Hence, the SM1 layer can bias the HM layer magnetization during switching, which significantly reduces the switching time distribution.

The presented situation would occur for pulsed HAMR. In this technique a laser spot moves over a recording medium and is just switched on at the desired write position, otherwise it is switched off. To ensure high data rates very short pulses with durations of about $100-200$\,ps are required. It was shown that for such ultrashort pulses DC noise is an issue for pure FePt like grains~\cite{vogler_areal_2016}, due to the short recording time window. The presented SM/FeRh/HM structure could solve this problem due to its reduced switching time distribution. Due to the low heat pulse temperatures, which are lower than the Curie temperature of all involved materials, all kinds of noise originating from thermal effects could be significantly reduced.

\section{Conclusion}
\label{sec:conclusion}
In this theoretical work we extensively investigated the benefits and problems of an exchange coupled high/low $T_{\mathrm{C}}$ grain structure with an FeRh interlayer for heat-assisted magnetic recording (HAMR). The FeRh layer, having a first order phase transition from an antiferromagnetic phase at room temperature to a ferromagnetic phase at high temperatures, tunes the exchange coupling between a soft magnetic (SM) high $T_{\mathrm{C}}$ layer and a hard magnetic (HM) low $T_{\mathrm{C}}$ one. The coupling has its full strength at high temperatures and at room temperature the exchange interactions reach their minimum. It is assumed that there remains a net ferromagnetic moment in the FeRh layer even at low temperatures, which is large enough to stabilize the magnetization of the SM layer along the direction of the HM layer magnetization, in order to achieve a high read-back signal. We calculated footprints of a typical HAMR head on various grain types, with a coarse grained Landau-Lifshitz-Bloch (LLB) model. These footprints point out that the transition jitter of the proposed SM/FeRh/HM structures is significantly reduced in both down-track direction by $20-30$\,\% and in off-track direction by $30-50$\,\%, compared to ordinary SM/HM grains. Additionally, due to the exchange spring effect no DC noise occurs even at high head velocities and high write temperatures, in contrast to pure FePt grains. The origin for the transition jitter reduction was found to be a fast increasing coercivity, or fast increasing freezing temperature, during cooling of the SM/FeRh/HM structure.

Further, a significant reduction of the switching time distribution of SM/FeRh/HM grains, by a factor of four, was found if ultrashort heat pulses, shorter than the write field duration are used. This can be obtained if the maximum temperature of the heat pulse is smaller than the Curie temperature of the involved materials. The decrease of the switching time distribution is based on a decoupling of the SM layer and the HM layer at room temperature, at which the FeRh interlayer has a small or no net magnetic moment. Hence, the SM layer can independently switch in write direction, and thus can bias the HM layer at higher temperatures. This mechanism narrows the distribution of the switching times, compared to ordinary SM/HM grains. 

The proposed grain with an FeRh interlayer in the middle of the SM/HM structure seems to combine the best of pure HM and ordinary SM/HM grains. However, all presented results assume that a sufficiently thin FeRh layer can be produced, which meets two basic requirements i) a net magnetic moment at room temperature, which is large enough to stabilize the SM layer, but which is low enough that a pronounced phase transition is preserved and ii) a critical temperature of the phase transition which can be adapted. Individually, both properties seem to be physically possible, but it still requires some research effort to achieve both properties at the same time. This work demonstrates the benefits and chances of such a SM/FeRh/HM structure. If requirement i) is not obtainable, the presented results can still be instructive to explain the mechanisms how an FeRh capping layer in FeRh/HM grains, as proposed in Ref.~\cite{thiele_ferh/fept_2003}, can influence magnetic noise, compared to pure FePt grains.

\section{Acknowledgements}
The authors would like to thank the Vienna Science and Technology Fund (WWTF) under grant No. MA14-044, the Advanced Storage Technology Consortium (ASTC), and the Austrian Science Fund (FWF) under grant No. I2214-N20 for financial support. The computational results presented have been achieved using the Vienna Scientific Cluster (VSC).

\bibliography{/home/christoph/Dropbox/HAMR}

\end{document}